\documentclass[conference,letterpaper]{IEEEtran}

\newtheorem{theorem}{\bf{Theorem}}
\newtheorem{definition}{\bf{Definition}}

\newtheorem{remark}{Remark}                             
\usepackage{graphicx}
\usepackage{epsfig}
\usepackage{latexsym}
\usepackage{amsfonts}
\usepackage{here}
\usepackage{rawfonts}
\usepackage[latin1]{inputenc}
\usepackage[T1]{fontenc}
\usepackage{calc}
\usepackage{url}
\usepackage{enumerate}
\usepackage{color}
\usepackage[tbtags]{amsmath}
\usepackage{amssymb}
\usepackage{upref}
\usepackage{epic,eepic}
\usepackage{times}
\usepackage{dsfont}
\usepackage{comment}
\usepackage{cite}

\usepackage{graphicx}
\usepackage[font={small}]{caption}
\usepackage[font={small}]{subcaption}

\usepackage{stfloats}
\usepackage{mathtools}
\usepackage{amsmath}

\usepackage{algorithm}
\usepackage{algpseudocode}

\begin{document}
\title{Federated Learning via Lattice Joint Source-Channel Coding} 


\author{%
  \IEEEauthorblockN{Seyed~Mohammad~Azimi-Abarghouyi}
  \IEEEauthorblockA{
                    KTH Royal Institute of Technology \\
                    Stockholm, Sweden\\
                    Email: seyaa@kth.se}
  \and
  \IEEEauthorblockN{Lav~R. Varshney}
  \IEEEauthorblockA{
                    University of Illinois Urbana-Champaign\\
                    Urbana, IL, USA \\
                    Email: varshney@illinois.edu}
	
}


\maketitle


\begin{abstract}
	This paper introduces a universal federated learning framework that enables over-the-air computation via digital communications, using a new joint source-channel coding scheme. Without relying on channel state information at devices, this scheme employs lattice codes to both quantize model parameters and exploit interference from the devices. A novel two-layer receiver structure at the server is designed to reliably decode an integer combination of the quantized model parameters as a lattice point for the purpose of aggregation. Numerical experiments validate the effectiveness of the proposed scheme. Even with the challenges posed by channel conditions and device heterogeneity, the proposed scheme markedly surpasses other over-the-air FL strategies.
	
\end{abstract}
\vspace{-15pt}
\section{Introduction}
Given the rise of advanced wireless edge devices, federated learning (FL) addresses challenges like data privacy, latency, and bandwidth by enabling local machine learning processing, ensuring data remains on-device \cite{mcmahan}. In IoT-dense wireless settings with network constraints, FL faces communication issues. While FL methods using orthogonal multiple access transmission prevent interference, they result in latency and require more resources \cite{pooor}. To address these challenges, a \textit{source coding scheme} based on lattice codes has been used in \cite{eldar, eldar2} to quantize model parameters in FL. In \cite{eldar2}, privacy enhancement is also considered. This use of lattice-based quantization is based on the principles of subtractive dithered lattice quantization \cite{rzamir1}, which is grounded in information-theoretic arguments. The method in \cite{eldar} offers a more precise finite-bit representation for FL than traditional scalar quantization methods. In a sense, it uses functional quantization \cite{varsh2}.

Leveraging interference from simultaneous edge device transmissions, over-the-air computation \cite{nazer, varsh} offers another strategy. Over-the-air FL, built on this, manages FL aggregation amidst interference, making it more efficient, resource-conservative, and faster than FL using orthogonal transmission. However, previous research on over-the-air FL has assumed analog modulation, where the transmitter can freely shape the carrier waveform by opting for any real number I/Q coefficients \cite{huang_analog, ding, gunduz2, cao, gunduz3}. This assumption may not be possible for existing wireless devices, as they come with digital modulation chips that may not support arbitrary modulation schemes. Moreover, following a power control approach, past work requires perfect channel state information at the transmitter (CSIT) for all devices to determine their transmission powers to counteract wireless channel effects during the aggregation process. In this approach, there exist physical constraints on both instantaneous and average transmission power. Consequently, a device experiencing poor channel conditions may either be unable to engage in the learning process or be required to utilize exceptionally high power \cite{huang_analog,ding, gunduz2, gunduz3, cao}. Past work also requires perfect synchronization among the transmitters. These prerequisites lead to a substantial overhead for channel estimation training and feedback before any transmission, causing increased delays and reduced spectral efficiency.

The special case of over-the-air FL using BPSK modulation is studied in \cite{gunduz3}. While significantly reducing implementation cost and resource requirements, one-bit quantization can result in high latency and performance gap with the non-quantized case \cite{onebit}. Furthermore, other challenges related to channel estimation and power control continue to persist. 

Most wireless systems use a constant power for blind transmission, not directly adjusting for the channel. Beyond not requiring CSIT and power control, this approach can offer a range of advantages. Firstly, it enables the maintenance of average energy for signal transmission, no matter how the channel varies. Additionally, it helps prevent the dynamic range of the transmitted signal from enlarging, making hardware implementations significantly simpler and reducing costs. Lastly, approaches that adjust for the channel at the transmitter can run into channel estimation errors, causing the values to be multiplied by unpredictable gains when received \cite{cohen1}. Considering these, \cite{gunduz4} proposes a blind over-the-air FL strategy. However, by suggesting a particular receiver equalization, this method calls for a large number of receiver antennas, and the impact of wireless fading diminishes as the antenna count grows without bound. On the other hand, \cite{cohen1, cohen2} investigate blind over-the-air FL that does not compensate for the adverse effects of fading. 


In this work, we propose a \textit{joint source-channel coding scheme} that incorporates novel transmission and aggregation strategies using lattice codes with adjustable quantization levels. Developing a novel coding base, our aim is to enhance the resilience of over-the-air FL against noise and interference, thus ensuring the realization of desired learning outcomes. This particularly ensures effective performance in challenging channel conditions, even with a limited number of antennas available at the server. Notably our blind approach does not rely on any prior knowledge or CSIT. It is inspired by past research \cite{nazer_cmp, azimi_cf1, azimi_cf2, azimi_cf3} which indicates advantages of nested lattice coding based modulation in compute-and-forward relaying as a \textit{channel coding scheme}, enhancing bit-rate throughput at relays. The key contributions are as follows.

\textit{Compute-Update Scheme:} We provide an end-to-end real-valued model parameter transmission framework for FL, named compute-update FL--- {\fontfamily{lmtt}\selectfont
	FedCPU}. On the transmitter side, we develop lattice codes for quantization of the model parameters. Our quantization approach is based on normalization and dithering. On the server side, we decode a single integer combination of transmitted quantized model parameters as a lattice point, which  after processing yields a new form of aggregation. 
The reason for using integer combinations is because the aggregation of model parameters in FL has an additive structure. Additionally, lattice Voronoi region over the decoded integer combination protects against the decoding error arising from interference and noise. This level of protection is absent in analog modulation schemes, given their continuous transmitted values \cite{huang_analog,ding, gunduz2, gunduz3, cao, cohen1, cohen2, gunduz4}. Also, the adjustable aggregation method we propose marks a significant shift from the traditional practice in FL of employing fixed, predefined weights, which are either equal or based on the size of local datasets. This conventional approach, recommended for ideal, error-free communication conditions as outlined in the seminal FL paper \cite{mcmahan}, has been consistently used in over-the-air FL research \cite{huang_analog, ding, mcmahan, pooor, gunduz2, gunduz3, cao, cohen1, cohen2, gunduz4, onebit, eldar, eldar2}. However, this persists despite the fact that over-the-air FL deals with imperfect communication scenarios, plagued by interference and noise. In response to this, our method strategically tailor the aggregation weights to address the wide range of decoding errors unique to each integer combination in our model.

\textit{Transmitter-Receiver Architecture:} To accomplish the aggregation, we propose a two-layer receiver architecture at the server side that incorporates both the real and imaginary parts of the signal. The initial layer incorporates an equalization vector, which is subsequently optimized to minimize the decoding error. In the subsequent layer, a normalizing factor is introduced and optimized to minimize the quantization error.

\textit{System Insights:} Our experimental results highlight the efficacy of {\fontfamily{lmtt}\selectfont
	FedCPU} in addressing the challenges posed by the lack of CSIT and the presence of only a limited number of antennas at the server. Shrinking the lattice Voronoi regions reveals a balance between increased decoding error and diminished quantization error. Notably, {\fontfamily{lmtt}\selectfont
	FedCPU} showcases markedly superior learning accuracy compared to other over-the-air schemes, due to its distinctive aggregation technique and finely-tuned receiver design. The learning results closely follow those expected in an ideal, error-free orthogonal transmission.

\vspace{0pt}

\section{System Model}
\vspace{-5pt}
\subsection{Setup}
There are $K$ devices and a single server as the basic setup for FL systems, see Fig. 1. All the devices are single-antenna units, but the server has $M$ antennas. The downlink channels from the server to the devices are considered error-free. The uplink channel from each device $k$ to the $m$th antenna of the server at communication round $t$ is represented by ${h}_{mk,t} \in {\mathbb{C}}$. All devices are considered to have the same transmission power constraint $P$. However, by appropriately adjusting the channel coefficients, we can integrate asymmetric power constraints. Let the entire channel matrix $\mathbf{H}_{t}^{\text{c}} \in {\mathbb{C}}^{M \times K}$ be 
\begin{align}
\small
\mathbf{H}_{t}^{\text{c}} =
\begin{bmatrix}
{h}_{11,t} & \cdots & {h}_{1K,t} \\
\vdots & \ddots & \vdots \\
{h}_{M1,t} & \cdots & {h}_{MK,t} \\
\end{bmatrix}.
\end{align}
The server is the only node that knows $\mathbf{H}_t^{\text{c}}$ as the channel state information at the receiver (CSIR) \textit{after the transmission}, whereas the devices have no knowledge of channels. 

\subsection{Learning Algorithm}
A frequently adopted FL algorithm named {\fontfamily{lmtt}\selectfont FedAvg} \cite{mcmahan} is elucidated as follows. For a specific round $t$ within the set $\{0,\ldots,T-1\}$, where $T$ is the total number of rounds, every device $k$ initiates by refining its individual learning model over $\tau$ local training epochs. Each of these epochs utilizes a mini-batch $\boldsymbol\xi_k^i$ with size of $B$ selected at random from the local dataset $\mathcal{D}_k$. This process is further detailed as $
\mathbf{w}_{k,t,i+1} = {\mathbf{w}_{k,t,i}}- \mu_t \nabla F_k(\mathbf{w}_{k,t,i}, \boldsymbol\xi_k^i), \forall i \in \left\{0,\ldots,\tau-1\right\}$,
where $\mu_t$ is the learning rate at round $t$, and $\nabla F_k(\mathbf{w}_{k,t,i}, \boldsymbol\xi_k^i)$ is the local gradient computed at the local model parameter vector $\mathbf{w}_{k,t,i}$. Then, each device $k$ uploads the local model update $\Delta\mathbf{w}_{k,t} = {\mathbf{w}_{k,t,\tau}} - {\mathbf{w}_{k,t,0}}$ to the server for aggregation. For ideal aggregation, the global model update can be achieved by averaging the model updates from all devices, giving them equal weights, as
\begin{align}
\label{agg}
\Delta\mathbf{w}_{\text{G},t+1} = \mathbf{w}_{\text{G},t+1} - \mathbf{w}_{\text{G},t} = \frac{1}{K}\sum_{k=1}^{K} \Delta\mathbf{w}_{k,t}.
\end{align}
Subsequently, the server broadcasts the acquired global model $\mathbf{w}_{\text{G},t+1}$ to all devices. Using this model, each device $k$ sets its starting state for the upcoming round as $\mathbf{w}_{k,t+1,0} = \mathbf{w}_{\text{G},t+1}$. 

\vspace{-5pt}
\subsection{Lattice Preliminaries}
\begin{definition}
	A lattice $\Lambda$ is a discrete subgroup of $\mathbb{R}^{s \times 1}$ and can be expressed as a linear transformation of integer vectors as $
	\Lambda = \left\{\mathbf{G}\mathbf{s}: \mathbf{s} \in \mathbb{Z}^{s \times 1}\right\}$,
	where $\mathbf{G} \in \mathbb{R}^{s\times s}$ is the lattice generator matrix. In $\Lambda$, any linear integer combination of lattice points is itself a lattice point. Lattice codebooks can be formed by combining a regular (i.e., low-dimensional lattice) constellation
	(e.g., PAM, QAM) with a linear code (e.g., LDPC
	code) or via more intricate nested lattice constructions \cite{erez}. 
\end{definition}
\begin{definition}
	A lattice quantizer is a map $Q_{\Lambda}: \mathbb{R}^{s \times 1} \to \Lambda$ that quantizes a point to its nearest lattice point based on
	Euclidean distance as $
	Q_{\Lambda} (\mathbf{x}) = \arg\min_{\lambda \in \Lambda} \|\mathbf{x}-\lambda\|^2$.
\end{definition}

\begin{definition}
	Fundamental Voronoi region $\mathcal{V}$ is the set of all
	points that are quantized to the origin as
	$\mathcal{V} = \left\{\mathbf{x} \in \mathbb{R}^{s \times 1} : Q_{\Lambda}(\mathbf{x}) = \mathbf{0}\right\}$.
\end{definition}

\begin{definition}
	The second moment of a lattice is the per-dimension second moment of a uniform distribution over the foundational Voronoi region $\cal V$ as $\sigma_\text{q}^2 = \frac{\int_{\mathcal{V}}^{}\|\mathbf{x}\|^2\mathrm{d}\mathbf{x}}{s\int_{\mathcal{V}}^{}\mathrm{d}\mathbf{x}}$.
	
\end{definition}

\vspace{0pt}

\section{FedCPU: Compute-Update Scheme}
{\fontfamily{lmtt}\selectfont
	FedCPU} consists of two primary components: the transmission scheme employed by the devices and the aggregation scheme used by the server. The {\fontfamily{lmtt}\selectfont FedCPU} develops a new version of the aggregation vector. This version features adjustable aggregation weights through integer coefficients, making use of the lattice structure and the additive properties of wireless multiple-access channels. Unlike over-the-air FL schemes in \cite{huang_analog,ding, gunduz2, gunduz3, cao} that depend on CSIT, {\fontfamily{lmtt}\selectfont FedCPU} operates with constant power and does not rely on prior information. This sets {\fontfamily{lmtt}\selectfont FedCPU} apart from these schemes that use power control, imposing average and maximum power constraints, and necessitate device selection. With {\fontfamily{lmtt}\selectfont FedCPU}, every device participates in the learning process based on an appropriate aggregation weight tailored to its communication conditions, free from the mentioned constraints. Additionally, though Sec.\ II.B discusses a specific algorithm, {\fontfamily{lmtt}\selectfont FedCPU} is not limited to it. For simplicity, the iteration index is omitted here.
\vspace{0pt}
\subsection{Transmission Scheme} Based on a lattice $\Lambda$, the transmission preparation procedure carried out by each device $k$ is as follows.

1) The model update parameters undergo normalization to achieve zero mean and unit variance. The normalized model update is obtained as ${\widehat{\Delta\mathbf{w}}_{k}} = \frac{{\Delta{\mathbf{w}}_k}-\vartheta_k\mathbf{1}}{\sigma_k}$, where $\mathbf{1}$ is the all-one vector, and $\vartheta_k$ and ${\sigma_k}$ denote the mean and standard deviation of the $s$ entries of the model update vector given by $
\vartheta_k = \frac{1}{s}\sum_{i=1}^{s} \delta w_{k,i},\
\sigma_k^2 = \frac{1}{s}\sum_{i=1}^{s}(\delta w_{k,i}-\vartheta_k)^2$,
where $\delta w_{k,i}$ is the $i$-th entry of the vector $\Delta \mathbf{w}_k$. Each device $k$ shares the two scalars $(\vartheta_k,\sigma_k)$ error-free to the server via an orthogonal feedback channel.

2) A dither vector $\mathbf{d}_{k} \in \mathcal{V}$
is generated independently from other devices according to a random uniform distribution.

3) The dither is added to $\widehat{\Delta\mathbf{w}}_k$ and the nearest codeword to the result is selected as
\begin{align}
{\overline {\Delta\mathbf{w}}_{k}} = Q_{\Lambda}( \widehat {\Delta\mathbf{w}}_k+\mathbf{d}_{k}).
\end{align}
The dither addition results in the quantization error becoming uniform. For the quantization error $\boldsymbol\epsilon_k$ defined as $Q_{\Lambda}( \widehat {\Delta\mathbf{w}}_k+\mathbf{d}_{k})-\mathbf{d}_k = \widehat{\Delta \mathbf{w}_k} + \boldsymbol\epsilon_k$, it holds that $\mathbb{E}\left\{\|\boldsymbol\epsilon_k\|^2\right\} = \sigma_\text{q}^2$.

4) The device scales the resulting lattice point and transmits
\begin{align}
\mathbf{x}_{k} =  \sqrt{\frac{P}{1+2\sigma_\text{q}^2}} {\overline {\Delta\mathbf{w}}_{k}},
\end{align}
whereby we have $
\mathbb{E}\left\{\|\mathbf{x}_k\|^2\right\} = \frac{P}{1+2\sigma_\text{q}^2}\mathbb{E}\left\{\|{\overline {\Delta\mathbf{w}}_{k}}\|^2\right\} \leq \frac{P}{1+2\sigma_\text{q}^2} \left(\mathbb{E}\left\{\|\widehat {\Delta\mathbf{w}}_k+\mathbf{d}_{k}\|^2\right\} + \sigma_\text{q}^2\right)= \frac{P}{1+2\sigma_\text{q}^2}\times \left(\mathbb{E}\left\{\|\widehat {\Delta\mathbf{w}}_k\|^2\right\}+\mathbb{E}\left\{\|\mathbf{d}_{k}\|^2\right\} + \sigma_\text{q}^2\right) = P$,
which satisfies the power constraint at the devices.

The server knows all dither vectors due to shared randomness with the devices, a common assumption in literature \cite{eldar, eldar2,rzamir1,nazer_cmp, azimi_cf1, azimi_cf2, azimi_cf3, erez}. Fig. 2
provides a schematic representation of the transmitter architecture.
\begin{figure}[tb!]
	\centering
	
	\includegraphics[width =2.7in]{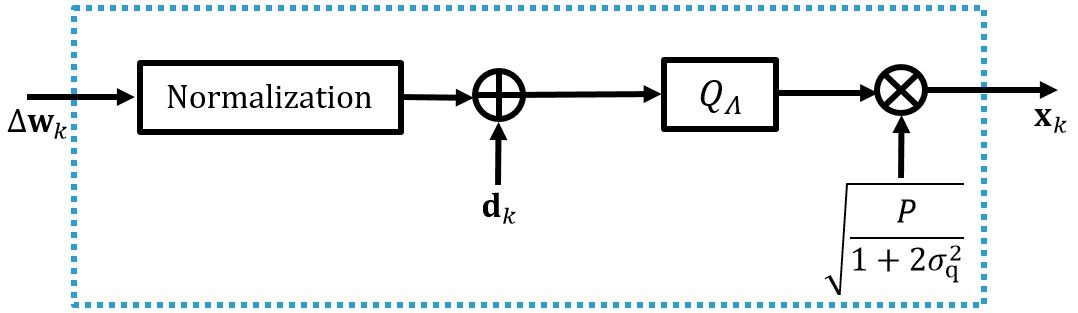}
	
	\caption{Transmitter structure for device $k$.}
	\vspace{-15pt}
\end{figure}
\subsection{Aggregation Scheme} 
The baseband received signal at antenna $m$ of the server, $\mathbf{y}_m \in \mathbb{C}^{{s}\times 1}$, is $\mathbf{y}_m = \sum_{k=1}^{K}h_{mk} \mathbf{x}_k + \mathbf{z}_m$, where $\mathbf{z}_m \in \mathbb{C}^{{s}\times 1}$ is complex Gaussian noise, with each entry having a variance of $\sigma_\text{z}^2$. 
Thus, the real-valued representation is as follows.
\begin{align}
\label{realsignal}
\mathbf{Y} = {\mathbf{H}}\mathbf{X} + \mathbf{Z},
\end{align}
where $\mathbf{Y} = \left[\mathfrak{Re}\left\{\mathbf{y}_1\right\}, \ldots,\mathfrak{Re}\left\{\mathbf{y}_M\right\},
\mathfrak{Im}\left\{\mathbf{y}_1\right\}, \ldots, \mathfrak{Im}\left\{\mathbf{y}_M\right\}\right]^\top$, ${\mathbf{H}} = \begin{bmatrix}
\mathfrak{Re}\left\{\mathbf{H}^\text{c}\right\}  \\
\mathfrak{Im}\left\{\mathbf{H}^\text{c}\right\}
\end{bmatrix}$, $\mathbf{X} = [\mathbf{x}_1, \ldots, \mathbf{x}_K]^\top$, $\mathbf{Z} = [
\mathfrak{Re}\left\{\mathbf{z}_1\right\} , \ldots ,\mathfrak{Re}\left\{\mathbf{z}_M\right\},
\mathfrak{Im}\left\{\mathbf{z}_1\right\},\ldots, \mathfrak{Im}\left\{\mathbf{z}_M\right\}]^\top$.
Based on \eqref{realsignal} and the transmission scheme in Sec.\ III.A, the receiver architecture for the aggregation at the server is as follows.

1) An equalization vector $\mathbf{b} \in \mathbb{R}^{2 M \times 1}$ is used to obtain
\begin{align}
\mathbf{b}^\top \mathbf{Y} = \mathbf{b}^\top {\mathbf{H}} \mathbf{X}+ \mathbf{b}^\top \mathbf{Z}.
\end{align}

2) The decoder aims to recover a lattice point or equally a linear integer combination of quantized model updates $\sum_{k=1}^{K}a_k \overline{\Delta\mathbf{w}}_k^\top = \mathbf{a}^\top \overline{\Delta\mathbf{W}}$ directly from $\mathbf{b}^\top \mathbf{Y}$, where $\mathbf{a} = [a_1,\ldots,a_{K}]^\top \in \mathbb{Z}^{K \times 1}$ denotes the integer coefficient vector and $
\overline{\Delta\mathbf{W}} =
[\overline{\Delta\mathbf{w}}_{1},
\ldots,
\overline{\Delta\mathbf{w}}_{K}]^\top$.

For this decoding, the result $\mathbf{b}^\top \mathbf{Y}$ is scaled and then quantized to its nearest lattice point as
\begin{align}
Q_{\Lambda}\Biggl(\sqrt{\frac{1+2\sigma_\text{q}^2}{P}}\mathbf{b}^\top \mathbf{Y}\Biggr) = \mathbf{a}^\top \overline{\Delta\mathbf{W}}.
\end{align}

\begin{figure}[tb!]
	\centering
	
	\includegraphics[width =3in]{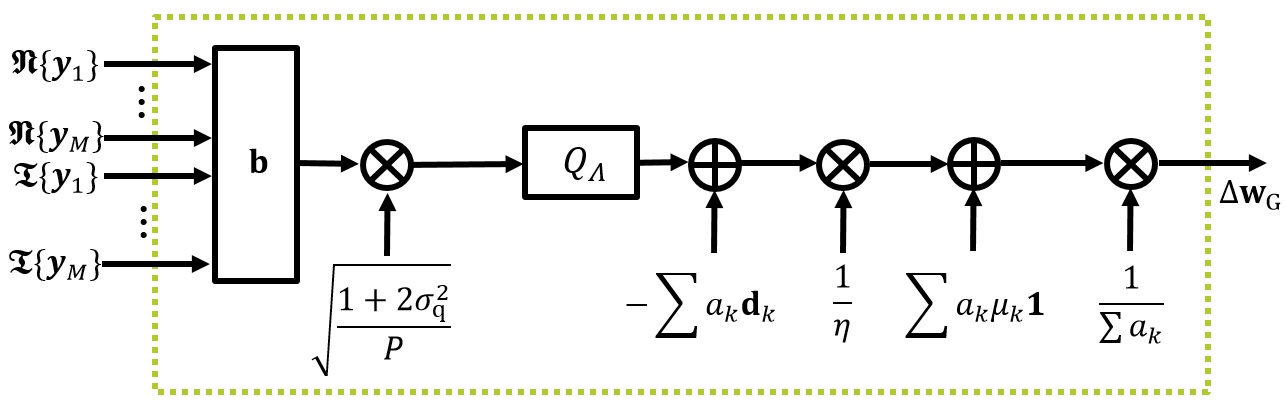}
	
	\caption{Receiver structure for the server.}
	\vspace{-15pt}
\end{figure}

We can rewrite $\sqrt{\frac{1+2\sigma_\text{q}^2}{P}}\mathbf{b}^\top \mathbf{Y}$ as $\mathbf{a}^\top \overline{\Delta\mathbf{W}}+\sqrt{\frac{1+2\sigma_\text{q}^2}{P}}(\mathbf{b}^\top{\mathbf{H}}-\mathbf{a}^\top)\mathbf{X}+\sqrt{\frac{1+2\sigma_\text{q}^2}{P}}\mathbf{b}^\top \mathbf{Z}$.
Thus, the decoding error in recovering $\mathbf{a}^\top \overline{\Delta\mathbf{W}}$ is measured in terms of the mean squared error (MSE), referred to as decoding MSE, as follows.
\begin{align}
\label{mse}
&\text{DMSE}(\mathbf{a}) = \mathbb{E}\Biggl\{\Biggl\Vert \sqrt{\frac{1+2\sigma_\text{q}^2}{P}}(\mathbf{b}^\top{\mathbf{H}}-\mathbf{a}^\top)\mathbf{X}+\sqrt{\frac{1+2\sigma_\text{q}^2}{P}}\times\nonumber\\
&\mathbf{b}^\top \mathbf{Z}\Biggr\Vert^2\Biggr\} =({1+2\sigma_\text{q}^2})\left(\|\mathbf{b}^\top{\mathbf{H}}-\mathbf{a}^\top\|^2+\frac{1}{\text{SNR}}\|\mathbf{b}\|^2\right),
\end{align}
where the independence of $\mathbf{x}_k$ and $\mathbf{x}_{k'}$, $\forall k\neq k'$, is assumed, due to the transmit normalization and dithering. Independence is also a factor in other over-the-air FL schemes \cite{ding, cao}. Also, $\text{SNR} = \frac{P}{\sigma_\text{z}^2}$ denotes the signal-to-noise ratio. The next theorem presents the vector $\mathbf{b}$ that minimizes the decoding MSE.  
\begin{theorem}
	The optimal equalization vector for a given coefficient vector $\mathbf{a}$ is
	\begin{align}
	\label{bopt}
	\mathbf{b}_\text{opt}^\top = \mathbf{a}^\top {\mathbf{H}}^\top \left(\frac{1}{\text{SNR}}\mathbf{I}+{\mathbf{H}}{\mathbf{H}}^\top\right)^{-1}.
	\end{align}
\end{theorem}
\begin{IEEEproof}
	Expanding $\frac{\text{DMSE}(\mathbf{a})}{{1+2\sigma_\text{q}^2}}$ as
	\begin{align}
	&\|\mathbf{b}^\top{\mathbf{H}}-\mathbf{a}^\top\|^2+\frac{1}{\text{SNR}}\|\mathbf{b}\|^2 = \left(\mathbf{b}^\top{\mathbf{H}}-\mathbf{a}^\top\right)\left({\mathbf{H}}^\top\mathbf{b}-\mathbf{a}\right)+\nonumber\\
	&\hspace{-4pt}\frac{1}{\text{SNR}}\|\mathbf{b}\|^2 = \mathbf{b}^\top{\mathbf{H}}{\mathbf{H}}^\top\mathbf{b} -2\mathbf{b}^\top{\mathbf{H}}\mathbf{a}+\mathbf{a}^\top\mathbf{a}+\frac{1}{\text{SNR}}\mathbf{b}^\top\mathbf{b}
	\end{align}
	and taking derivative from the result with respect to $\mathbf{b}$, we obtain $2{\mathbf{H}}{\mathbf{H}}^\top \mathbf{b} - 2 {\mathbf{H}}\mathbf{a}+\frac{2}{\text{SNR}}\mathbf{b}$,
	which amounts to zero at \eqref{bopt}.
\end{IEEEproof}
Substituting $\mathbf{b}_\text{opt}$ into \eqref{mse}, we obtain
\begin{align}
\label{dmse_mil}
&\text{DMSE}(\mathbf{a}) = ({1+2\sigma_\text{q}^2})\times \nonumber\\&\mathbf{a}^\top\left[ \mathbf{I} - {\mathbf{H}}^\top \left(\frac{1}{\text{SNR}}\mathbf{I}+{\mathbf{H}}{\mathbf{H}}^\top\right)^{-1}{\mathbf{H}}\right]\mathbf{a}.
\end{align}
Applying the matrix inversion lemma \cite{matrix_inversion}, \eqref{dmse_mil} can be expressed as
\begin{align}
\label{dmsee}
\text{DMSE}(\mathbf{a}) = ({1+2\sigma_\text{q}^2})\mathbf{a}^\top\left( \mathbf{I} +{\text{SNR}}{\mathbf{H}}^\top{\mathbf{H}}\right)^{-1}\mathbf{a}.
\end{align}
From the $\text{SNR}$ and $\mathbf{H}$ as seen in \eqref{dmsee}, the DMSE arises due to interference and noise impacting the decoding process. 

The value of $\text{DMSE}(\mathbf{a})$ can be interpreted as the variance of the effective noise in decoding $\mathbf{a}^\top \overline{\Delta\mathbf{W}}$. Therefore, the decoding error probability, which represents the probability of decoding any other lattice point besides $\mathbf{a}^\top \overline{\Delta\mathbf{W}}$, is equated to the event where the decoding noise lies outside $\mathbf{a}^\top \overline{\Delta\mathbf{W}}+\mathcal{V}$.
\begin{remark}
	For a decoding noise within $\mathbf{a}^\top \overline{\Delta\mathbf{W}}+\mathcal{V}$, the system is fully protected from interference and noise.
\end{remark}

\begin{remark}
	Unlike conventional digital communications, where a decoding error results in an entirely incorrect message, in {\fontfamily{lmtt}\selectfont
		FedCPU} a decoding error merely introduces an additive estimation error in the subsequent equation \eqref{decode_error_add}. This estimation error depends on the distance between the decoded lattice point and the target lattice point. In essence, while our goal is to minimize the decoding error to enhance system performance, it does not constitute a critical constraint or a bottleneck for the system. 
\end{remark}

3) After decoding and removing the dithers, the aggregation vector at the output of the receiver is estimated as
\begin{align}
\label{model_estimate}
\Delta\mathbf{w}_\text{G}^\top = \frac{Q_{\Lambda}\left(\sqrt{\frac{1+2\sigma_\text{q}^2}{P}}\mathbf{b}^\top \mathbf{Y}\right) - \mathbf{a}^\top {\mathbf{D}}}{\eta \mathbf{1}^\top\mathbf{a}}+\frac{1}{ \mathbf{1}^\top\mathbf{a}}\sum_{k=1}^{K} a_k\vartheta_k \mathbf{1}^\top,
\end{align}
where $\eta$ is a normalizing factor, the summation $\mathbf{1}^\top\mathbf{a}$ is to take the average for the aggregation, and $\mathbf{D} = [
{\mathbf{d}}_{1},
\ldots,
{\mathbf{d}}_{K}]^\top$.
Due to the transmit normalization, the estimation in \eqref{model_estimate} is unbiased, and one can consider any vector $\mathbf{a}$ with non-negative integer coefficients except the all-zero vector $\mathbf{0}$ for an aggregation. It differes from other over-the-air schemes, e.g., \cite{ding, pooor, mcmahan,huang_analog, gunduz2, gunduz3, cao, cohen1, cohen2, gunduz4, onebit, eldar, eldar2}, which use predefined weights in FL, such as equal or proportional to local dataset sizes.

We can rewrite \eqref{model_estimate} as
\begin{align}
\label{decode_error_add}
&\Delta\mathbf{w}_\text{G}^\top = \frac{1}{\mathbf{1}^\top \mathbf{a}} \sum_{k=1}^{K}a_k \Delta \mathbf{w}_k^\top + \frac{\mathbf{a}^\top\overline{\Delta\mathbf{W}}-\mathbf{a}^\top \mathbf{D}}{\eta \mathbf{1}^\top \mathbf{a}}+\nonumber\\
&\frac{\sum_{k=1}^{K} a_k \left(\vartheta_k\mathbf{1}^\top - \Delta \mathbf{w}_k^\top\right)}{ \mathbf{1}^\top\mathbf{a}} =\frac{1}{\mathbf{1}^\top \mathbf{a}} \sum_{k=1}^{K}a_k \Delta \mathbf{w}_k^\top +\nonumber\\
& \frac{\mathbf{a}^\top\overline{\Delta\mathbf{W}}-\mathbf{a}^\top \mathbf{D}}{\eta \mathbf{1}^\top \mathbf{a}}-\frac{\sum_{k=1}^{K}a_k\sigma_k\widehat{\Delta \mathbf{w}_k}^\top}{\mathbf{1}^\top \mathbf{a}} =\frac{1}{\mathbf{1}^\top \mathbf{a}} \sum_{k=1}^{K}a_k \Delta \mathbf{w}_k^\top \nonumber\\
&\hspace{-8pt}+ \frac{\sum_{k=1}^{K}a_k\widehat{\Delta\mathbf{w}_k}^\top+\sum_{k=1}^{K}a_k \boldsymbol\epsilon_{k}^\top}{\eta \mathbf{1}^\top \mathbf{a}}-\frac{\sum_{k=1}^{K}a_k\sigma_k\widehat{\Delta \mathbf{w}_k}^\top}{\mathbf{1}^\top \mathbf{a}}
\end{align}
In \eqref{decode_error_add}, the first term is the desirable aggregation and the remaining terms are mainly due to the quantization errors. Thus, the overall error against this estimation in terms of the MSE, referred to as quantization MSE, is
\begin{align}
&\text{QMSE}(\mathbf{a}) = \mathbb{E}\Biggl\{\Biggl\Vert\frac{\sum_{k=1}^{K}a_k\widehat{\Delta\mathbf{w}_k}+\sum_{k=1}^{K}a_k \boldsymbol\epsilon_{k}}{\eta \mathbf{1}^\top \mathbf{a}}-\frac{1}{\mathbf{1}^\top \mathbf{a}}\times\nonumber\\
&\sum_{k=1}^{K}a_k\sigma_k\widehat{\Delta \mathbf{w}_k}\Biggr\Vert^2\Biggr\}= \frac{1}{(\mathbf{1}^\top\mathbf{a})^2}\Biggl(\mathbb{E}\Biggl\{\Biggl\Vert \sum_{k=1}^{K} \left(\frac{a_k}{\eta} - a_k \sigma_k\right)\times\nonumber\\
&\widehat{\Delta\mathbf{w}_k} \Biggr\Vert^2\Biggr\}+\frac{1}{\eta^2}\mathbb{E}\left\{\left\Vert\sum_{k=1}^{K}a_k \boldsymbol\epsilon_{k}\right\Vert^2\right\}\Biggr) = \frac{1}{(\mathbf{1}^\top\mathbf{a})^2}\times\nonumber\\
&\left(\left\Vert\left(\frac{1}{\eta}\mathbf{I}-\text{diag}(\boldsymbol \sigma)\right)\mathbf{a}\right\Vert^2+\frac{1}{\eta^2}\|\mathbf{a}\|^2 \sigma_\text{q}^2\right) = \frac{1}{(\mathbf{1}^\top\mathbf{a})^2}\times\nonumber\\
&\left(\mathbf{a}^\top\left(\frac{1}{\eta}\mathbf{I}-\text{diag}(\boldsymbol \sigma)\right)^2\mathbf{a}+\frac{1}{\eta^2}\|\mathbf{a}\|^2 \sigma_\text{q}^2\right),
\end{align}
where $\boldsymbol\sigma = [\sigma_1,\ldots,\sigma_K]^\top$. The factor $\eta$ that minimizes the quantization MSE is presented in the next theorem.  
\begin{figure*}
	\begin{subfigure}{0.33\textwidth}
		\centering
		\includegraphics[width =2in]{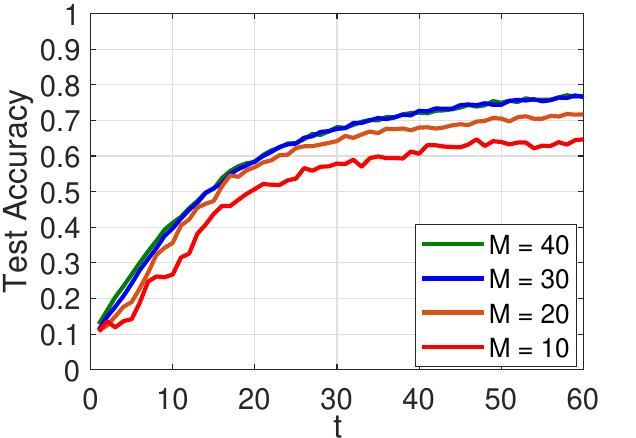} 
		\caption{}
	\end{subfigure}
	\begin{subfigure}{0.33\textwidth}
		\centering
		\includegraphics[width =2in]{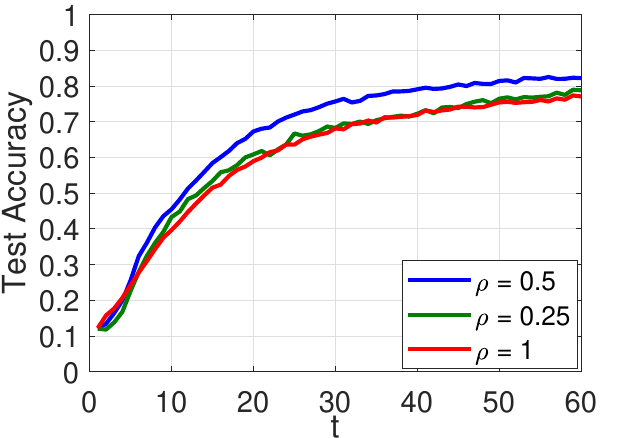} 
		\caption{}
	\end{subfigure}
	\begin{subfigure}{0.33\textwidth}
		\centering
		\includegraphics[width =2in]{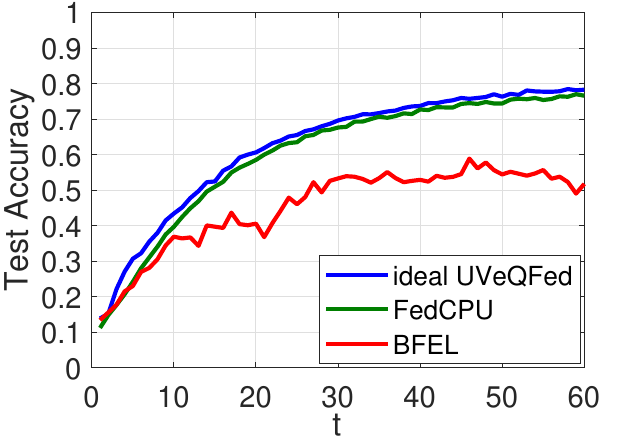} 
		\caption{}
	\end{subfigure}
	\vspace{-5pt}
	\caption{Test accuracy (a) for different $M$ values, (b) for different $\rho$ values, and (c) for different FL strategies.}
	\vspace{-15pt}
\end{figure*}
\begin{theorem}
	The optimal $\eta$ for a given $\mathbf{a}$ is
	\begin{align}
	\eta_\text{opt} = \frac{\left(1+\sigma_\text{q}^2\right)\|\mathbf{a}\|^2}{\mathbf{a}^\top \text{diag}(\boldsymbol \sigma)\mathbf{a}}.
	\end{align}
\end{theorem}
\begin{IEEEproof}
	We can expand $(\mathbf{1}^\top\mathbf{a})^2 \text{QMSE}(\mathbf{a})$ as
	\begin{align}
	\label{qmse}
	&\mathbf{a}^\top\left(\frac{1}{\eta}\mathbf{I}-\text{diag}(\boldsymbol \sigma)\right)^2\mathbf{a}+\frac{1}{\eta^2}\|\mathbf{a}\|^2 \sigma_\text{q}^2 =\nonumber\\
	& \mathbf{a}^\top\left(\frac{1}{\eta^2}\mathbf{I}-\frac{2}{\eta}\text{diag}(\boldsymbol{\sigma})+\text{diag}(\boldsymbol \sigma^2)\right)\mathbf{a}+\frac{1}{\eta^2}\|\mathbf{a}\|^2 \sigma_\text{q}^2.
	\end{align}	
	By taking derivative with respect to $\eta$ and equating the resulting expression to zero, we obtain
	\begin{align}
	&\mathbf{a}^\top\left(\frac{-2}{\eta^3}\mathbf{I}+\frac{2}{\eta^2}\text{diag}(\boldsymbol{\sigma})\right)\mathbf{a}-\frac{2}{\eta^3}\|\mathbf{a}\|^2 \sigma_\text{q}^2 =\nonumber
	\end{align}
	\begin{align}
	& \frac{1}{\eta} (1+\sigma_\text{q}^2)\|\mathbf{a}\|^2 - \mathbf{a}^\top \text{diag}(\boldsymbol \sigma) \mathbf{a}= 0,
	\end{align}	
	which leads to the final result.
\end{IEEEproof}
Substituting the optimal $\eta$, the QMSE is
\begin{align}
\label{errorQ}
&\text{QMSE}(\mathbf{a}) = \frac{1}{(\mathbf{1}^\top\mathbf{a})^2}\left(-\frac{1}{\eta_\text{opt}}\mathbf{a}^\top \text{diag}(\boldsymbol \sigma) \mathbf{a} +\mathbf{a}^\top \text{diag}(\boldsymbol \sigma^2) \mathbf{a}\right)\nonumber\\
&= \frac{1}{(\mathbf{1}^\top\mathbf{a})^2}\left(\mathbf{a}^\top \text{diag}(\boldsymbol \sigma^2) \mathbf{a} - \frac{\left(\mathbf{a}^\top \text{diag}(\boldsymbol \sigma) \mathbf{a}\right)^2}{(1+\sigma_\text{q}^2)\|\mathbf{a}\|^2}\right).
\end{align}
Fig. 3
provides a schematic representation of the receiver architecture
designed for this aggregation process. 

In {\fontfamily{lmtt}\selectfont
	FedCPU}, the coefficient vector $\mathbf{a}$ can be chosen using any metric to enhance learning performance. However, due to space limitations in presenting a learning convergence analysis of {\fontfamily{lmtt}\selectfont
	FedCPU}\footnote{A detailed convergence analysis of {\fontfamily{lmtt}\selectfont
		FedCPU} is presented in \cite{azimilav}, along with a corresponding metric which incorporates both DMSE and QMSE.}, we opt to minimize the decoding MSE as
\begin{align}
\arg\min_{\mathbf{a}\in \mathbb{Z}^{K\times 1}\backslash \mathbf{0}} \mathbf{a}^\top\left( \mathbf{I} +{\text{SNR}}{\mathbf{H}}^\top{\mathbf{H}}\right)^{-1}\mathbf{a}
\end{align}
subject to $a_k \geq 0, \ \forall k$. This optimization problem is an integer program that is NP-hard. Instead, we suggest the following suboptimal problem.
\begin{align} \hspace{-10pt}\text{round-to-integer}\left\{\arg\min_{\mathbf{a}\in \mathbb{R}^{K \times 1}} \mathbf{a}^\top\left( \mathbf{I} +{\text{SNR}}{\mathbf{H}}^\top{\mathbf{H}}\right)^{-1}\mathbf{a}\right\}
\end{align}
subject to $a_k \geq 1, \ \forall k$. This problem is convex since the matrix $\left( \mathbf{I} +{\text{SNR}}\mathbf{H}^\top\mathbf{H}\right)^{-1}$ is positive definite.

\section{Experimental Results}
The task classifies standard MNIST images (parameters as $K=30$, $\text{SNR} = 10$, $\tau = 3$, $\mu = 0.01$, $B = 100$, and $M = 30$) using a CNN. The CNN has two $3 \times 3$ convolutional layers with ReLU (32 and 64 channels), followed by $2 \times 2$ max pooling. It concludes with a 128-unit fully connected layer with ReLU and a softmax output. The lattice generator matrix is given by $\mathbf{G} = \text{diag}\left\{\mathbf{G}_2, \ldots, \mathbf{G}_2\right\}$, where $\mathbf{G}_2 = \begin{bmatrix}
0.25 & 0 \\
0.125 & 0.25 \\
\end{bmatrix}$, \cite{eldar, eldar2}. Devices have non-i.i.d. datasets with samples from two classes, varying in count per device. Performance, gauged by learning accuracy against global iteration count $t$, is averaged over 20 samples, considering a Gaussian channel with Rayleigh fading $\sim \exp(5)$ and uniform phase distribution $\sim {\cal{U}}(0,2{\pi})$.

In Fig. 3.(a), the accuracy is shown for different numbers of antennas $M$ at the server. The performance improves as $M$ increases because the decoding MSE is decreased. However, for higher values of $M$, this performance improvement tends to diminish. In Fig. 3.(b), we explore the implications of lattice quantization by employing different lattice generator matrices as 
$\rho \mathbf{G}$. As 
$\rho$ decreases, the lattice points become more densely packed, leading to smaller Voronoi regions. Notably, while a reduced 
$\rho$ minimizes the quantization errors, it does not necessarily enhance performance due to the rise in decoding error. This results in a balance of factors.

In Fig. 3.(c), comparisons are made between {\fontfamily{lmtt}\selectfont
	FedCPU} and schemes from the literature, serving as benchmarks. Among the evaluated benchmarks are {\fontfamily{lmtt}\selectfont ideal UVeQFed} \cite{eldar} and {\fontfamily{lmtt}\selectfont BFEL} \cite{gunduz4} schemes. The {\fontfamily{lmtt} \selectfont ideal UVeQFed} is the version of {\fontfamily{lmtt} \selectfont UVeQFed} when operating with ideal communications, given that {\fontfamily{lmtt} \selectfont UVeQFed} does not delve into the communication aspect of FL. It employs lattice quantization and leverages orthogonal transmission to completely avoid interference, operating under the premise of boundless communication resources. In contrast, the {\fontfamily{lmtt}\selectfont BFEL} scheme, which does not employ quantization, relies on multi-antenna processing at the server to mitigate interference effects. Notice {\fontfamily{lmtt}\selectfont
	FedCPU} significantly outperforms {\fontfamily{lmtt}\selectfont
	BFEL} scheme. The enhancement in performance is due to the use of adaptive aggregation weights, a factor previously overlooked. Additionally, we employ optimal equalization with a unique receiver architecture at the server, different from the equalization proposed in \cite{gunduz4}. Additionally, {\fontfamily{lmtt}\selectfont FedCPU}, subjected to interference and noise, nearly matches the performance of {\fontfamily{lmtt}\selectfont ideal UVeQFed}, which operates without interference and noise. 
\vspace{-5pt}
\section{Conclusions}
In this paper, we introduced a universal federated
learning scheme incorporating lattice codes, pioneering a new over-the-air computation method. The proposed scheme offers adjustable quantization, enabling distributed learning through digital modulation. Additionally, it ensures resilience against interference and noise through coding. In this scheme, with no need for channel compensation at the transmitter end, an integer combination of lattice quantized model parameters is reliably decoded and processed for aggregation. In spite of channel conditions and data heterogeneity, experimental findings showcased the superior learning accuracy of the proposed scheme, outperforming existing over-the-air alternatives. Furthermore, even with a limited number of antennas at the server, the proposed scheme can nearly achieve the performance level anticipated in a scenario devoid of errors.

\end{document}